\documentclass[aps,prd,twocolumn,superscriptaddress,floatfix,nofootinbib,longbibliography]{revtex4-2}

\usepackage{graphicx}
\usepackage{amsmath,amssymb,bm}
\usepackage{xcolor}
\usepackage[colorlinks=true,linkcolor=blue,citecolor=blue,urlcolor=blue]{hyperref}

\graphicspath{{figures/}}

\begin{document}

	\title{The Contribution of Mergers and Accretion to Dark Matter Halo Assembly:\texorpdfstring{\\}{ }An Excursion Set Theory Approach}

	\author{Hamed Kameli}
	\email{hkameli@gmail.com}
	\affiliation{Department of Physics, Center for Exact and Natural Sciences, Federal University of Paraíba, 58059-970, João Pessoa, Brazil}

	\begin{abstract}
We study the relative contributions of mergers and accretion to the assembly of dark matter haloes within the framework of Excursion Set Theory (EST). We extend a variance-discretized merger-tree algorithm, which recursively reconstructs all branches of a complete merger tree and introduces a merger-ratio criterion for distinguishing mergers from accretion. The method is validated by comparing the resulting progenitor distributions with the analytical conditional first up-crossing distribution, showing excellent agreement over a wide range of halo masses and redshifts. We find that the accretion contribution depends strongly on the adopted merger-ratio threshold. For large merger ratios ($MR=0.1$--$0.25$), accretion accounts for up to $\sim70\%$ of the mass growth of massive haloes, whereas including all resolved mergers ($MR=10^{-6}$) reduces the accretion contribution to $\sim40\%$. These values are in good agreement with previous N-body simulation results. The accreted mass exhibits a nearly self-similar distribution in the mass-redshift plane, with the dominant contribution arising from progenitors in the range $10^{-3}M_0 \lesssim M \lesssim 10^{-1}M_0$ and predominantly at $0.5 \lesssim z \lesssim 2$. Our results demonstrate that EST-based merger trees provide an accurate and computationally efficient framework for investigating halo assembly and quantifying the relative roles of mergers and accretion in hierarchical structure formation.		
	\end{abstract}

	\maketitle

	\section{Introduction}\label{sec:intro}
	The large-scale structure of the Universe emerges from the gravitational growth of primordial density fluctuations, which evolve into the cosmic web of voids, sheets, filaments, and clusters. As these fluctuations grow, matter collapses into gravitationally bound Dark Matter (DM) halos \cite{dodelson2003modern,peacock1999cosmological,padmanabhan1993structure,bernardeau2002large}. Dark matter halos subsequently grow through two principal processes: mergers, in which smaller halos combine to form larger halos, and accretion, the gradual incorporation of matter from their surroundings \cite{mo2010galaxy,somerville2015physical}. Understanding these processes is therefore essential for describing the formation and evolution of cosmic structures.\\
	 A statistical framework for predicting the abundance of collapsed objects was first developed by Press and Schechter \cite{Press1974} and later extended by Bond et al. through Excursion Set Theory (EST) \cite{Bond1991}. EST provides a stochastic description of density fluctuations across different smoothing scales and predicts the abundance and evolution of dark matter halos over cosmic time. In EST, the first up-crossing distribution determines the halo mass function and number density, while its conditional crossing yields the progenitor mass function \cite{Bond1991,Lacey1993}. The latter provides a statistical description of halo growth and serves as the basis for constructing merger trees and tracing assembly histories. The standard formulation assumes Markovian random walks and a spherical-collapse barrier, while more sophisticated extensions incorporate ellipsoidal collapse \cite{Sheth2001,Sheth2002} and non-Markovian effects \cite{maggiore2010halo,musso2012one,musso2012non,baghram2019exact,kameli2020modified,kameli2022mass}. \\ 
	Merger trees are commonly generated using discrete time-step algorithms, in which the merger history of a halo is reconstructed backward in redshift by evaluating the progenitor mass function at small time intervals. Using a uniform random number, the mass evolution is classified as either smooth accretion or a merger, and the procedure is recursively applied to all progenitor halos \cite{Cole2000,Neistein2008}. Several extensions have been proposed to incorporate non-binary mergers \cite{somerville1999plant}, calibrated result with N-body simulations \cite{Parkinson2008,cole2008statistical}, accretion probabilities \cite{alizadeh2010dark}, and ellipsoidal-collapse models  \cite{zhang2008conditional,Zhang2008,jiang2014generating}. An alternative approach, adopted in this work, constructs merger trees through discrete intervals in the variance $S$ rather than in time \cite{moreno2008merger,moreno2009dark}. This formulation is directly linked to the Markovian random-walk process underlying EST and has been shown to have a good agreement with N-body simulation results.\\
 	N-body simulations provide the most accurate numerical description of nonlinear structure formation and are frequently used to calibrate and validate analytical and semi-analytical models. Analyses of the Millennium simulation demonstrated that halo growth proceeds through a combination of mergers and diffuse accretion and provided widely used fitting functions for merger rates and mass assembly histories \cite{mcbride2009mass,Fakhouri2010}. Subsequent simulations, including Bolshoi, MultiDark, and IllustrisTNG, extended these studies over a broad range of halo masses and redshifts \cite{klypin2011dark,prada2012halo,klypin2016multidark,nelson2019illustristng,pillepich2018first}. Although direct measurements of halo merger and accretion rates remain difficult, and a detailed comparison with observations is beyond the scope of this work, observational studies of galaxy mergers, stellar halos, and Gaia substructures provide compelling evidence for hierarchical halo growth and assembly \cite{conselice2014galaxy,bell2017decoding,helmi2018merger,behroozi2022observational}.\\
 	Most previous studies have concentrated on merger rates and halo assembly histories rather than explicitly quantifying the relative contributions of mergers and accretion to halo growth. A pioneering study by Genel et al. \cite{genel2010growth}, based on merger trees extracted from the Millennium and Millennium-II simulations, found that resolved mergers account for approximately $60\%$ of halo mass growth, while the remaining $\sim40\%$ arises from smooth accretion of diffuse dark matter that was never previously bound to smaller halos. These results are in great agreement with our proposed results. \\
 	In this work, we develop an excursion-set merger-tree framework designed specifically to quantify the relative contributions of mergers and accretion to dark matter halo growth. Unlike most previous EST-based studies, which primarily focused on progenitor statistics and merger rates, our method follows the complete recursive branching structure of halo assembly and its sequences of accretion and merger. It employs an explicit merger-ratio criterion to distinguish merger from smooth accretion. We validate the results of merger tree with the analytical EST progenitor distributions and investigate how the merger and accretion contributions depend on halo mass, redshift, and merger-ratio threshold. The large statistical ensembles generated by the model further allow us to study a wide range of the mass and redshift distributions of the accreted component. Finally, a GPU implementation enables the construction of substantially larger merger-tree ensembles and improves the statistical precision of the results. \\
	 The paper is organized as follows. Section II reviews the theoretical framework of Excursion Set Theory and the concepts of halo assembly, mergers, and accretion. Section III presents the merger-tree algorithm and its implementation. Section IV provides validation tests and the main results on accretion contributions over a wide range of halo masses and redshifts. Section V summarizes our conclusions and discusses future directions.

	\section{Theoretical Foundation}\label{sec:theory}

\subsection{Excursion set formalism}\label{sec:est}
In this section, we briefly review the theoretical framework of Excursion Set Theory (EST), which provides a statistical description of the formation and abundance of non-linear cosmic structures. The foundations of this approach were established by Press \& Schechter \cite{Press1974}, who developed an analytical model for predicting the mass distribution of gravitationally collapsed objects. This formalism was later extended by Bond et al. \cite{Bond1991} into the excursion-set theory (EST) formalism. One of the major achievements of EST is its natural resolution of the cloud-in-cloud problem, thereby eliminating the fudge factor of two introduced in the original Press-Schechter derivation. By connecting the primordial linear density field to the non-linear population of dark matter (DM) halos, EST provides a powerful framework for predicting halo abundances, assembly histories, merger rates, and other statistical properties of structure formation. Comprehensive reviews can be found in \cite{zentner2007excursion} and Chapter~7 of \textit{Galaxy Formation and Evolution} \cite{mo2010galaxy}.\\
Within EST, the evolution of density perturbations is described by trajectories of the smoothed density contrast $\delta(S)$ in the $(S,\delta)$ plane, where $S(M)$ denotes the variance of the linear density field. Halo formation is assumed to occur when a trajectory first crosses a critical threshold, or collapse barrier, $\delta_c(z)$. For the spherical-collapse model, this threshold is given by
\begin{equation}
	\delta_c(z)=\frac{\delta_{c0}}{D(z)},
\end{equation}
where $\delta_{c0}=1.686$ is the critical overdensity for collapse at the present epoch and $D(z)$ is the linear growth factor normalized such that $D(0)=1$. The variance of the smoothed linear density field is defined as
\begin{equation}
	S(R)=\sigma^2(R)
	=
	\frac{1}{2\pi^2}
	\int_0^\infty dk\,k^2
	P_m(k,z=0)
	|\tilde W(kR)|^2,
\end{equation}
where $P_m(k,z)$ is the linear matter power spectrum and $\tilde W(kR)$ is the Fourier transform of the smoothing window function. Since larger smoothing scales suppress small-scale fluctuations, $S(R)$ decreases monotonically with increasing radius $R$ and the corresponding halo mass $M=(4\pi/3)\bar{\rho}_mR^3$. Consequently, large values of $S$ correspond to low-mass halos, while small values correspond to massive halos. The linear matter power spectrum is related to the primordial curvature perturbations through
\begin{equation}
	P_m(k,z)
	=
	\frac{8\pi^2}{25}
	\frac{k}{\Omega_M H_0^4}
	T^2(k)
	D^2(z)
	\mathcal{P}_{\mathcal{R}}(k),
\end{equation}
where $T(k)$ is the Eisenstein-Hu fitting formula for transfer function \cite{Eisenstein1998}. In the standard $\Lambda$CDM cosmology, the primordial curvature power spectrum is well approximated by
\begin{equation}
	\mathcal{P}_{\mathcal{R}}(k)
	=
	A_s
	\left(
	\frac{k}{k_p}
	\right)^{n_s-1},
\end{equation}
where $A_s$ is the amplitude of primordial fluctuations, $n_s$ is the spectral index, and $k_p$ is the pivot scale. Throughout this paper, we adopt the cosmological parameters reported by the Planck 2018 collaboration \cite{planck2020planck}.\\
A key feature of EST is the description of halo formation as a stochastic process. As the variance $S$ increases, the density contrast evolves as a random trajectory in the $(S,\delta)$ plane. For a sharp-$k$ filter these trajectories are Markovian, implying that successive steps are statistically independent. Halo formation is identified with the first up-crossing of the collapse barrier, and the corresponding crossing point variance uniquely determines the halo mass. The statistical properties of DM halos can therefore be obtained from the first up-crossing distribution of an ensemble of random walks.\\
For Markovian random walks and a constant spherical-collapse barrier, the first up-crossing distribution is given by \cite{Bond1991,zentner2007excursion}
\begin{equation}
	f_{\rm FU}(S,\delta_c)
	=
	\frac{1}{\sqrt{2\pi}}
	\frac{\delta_c}{S^{3/2}}
	\exp
	\left(
	-\frac{\delta_c^2}{2S}
	\right).
	\label{ffu}
\end{equation}
This quantity represents the fraction of trajectories whose first crossing occurs in the interval $(S,S+dS)$ and forms the basis for calculating the halo mass function. The number density of halos is
\begin{equation}
	\frac{dn}{dM}
	=
	\frac{\bar{\rho}_m}{M}
	f_{\rm FU}(S,\delta_c)
	\left|
	\frac{dS}{dM}
	\right|,
\end{equation}
which yields
\begin{equation}
	\frac{dn}{dM}
	=
	\frac{\bar{\rho}_m}{M}
	\frac{\delta_c(z)}
	{\sqrt{2\pi S^3}}
	\exp
	\left[
	-\frac{\delta_c^2(z)}
	{2S}
	\right]
	\left|
	\frac{dS}{dM}
	\right|.
\end{equation}
Beyond the unconditional simple mass function, EST can be extended to conditional mass functions, halo assembly histories, halo merger trees, progenitor mass distributions, and halo accretion histories. For Markovian trajectories, the conditional first up-crossing distribution can be derived analytically by shifting the origin of the random walk from $(0,0)$ to $(S_0,\delta_{c,0})$, corresponding to a host halo of mass $M_0$ identified at redshift $z_0$. The statistics of progenitor halos at an earlier epoch $z_1>z_0$ can then be obtained directly from the first up-crossing distribution evaluated relative to this new origin. Due to the translation invariance of Markovian walks, the conditional problem retains the same mathematical form as the unconditional one. The conditional first up-crossing distribution for spherical collapse model is therefore
\begin{equation}
	\begin{split}
		f(S_1,\delta_{c,1}|S_0,\delta_{c,0})\,dS_1
		&=
		\frac{1}{\sqrt{2\pi}}
		\frac{\delta_{c,1}-\delta_{c,0}}
		{(S_1-S_0)^{3/2}}
		\\
		&\times
		\exp
		\left[
		-\frac{(\delta_{c,1}-\delta_{c,0})^2}
		{2(S_1-S_0)}
		\right]
		dS_1,
	\end{split}
	\label{eq:conditional_EST}
\end{equation}
where $S_i\equiv S(M_i)=\sigma^2(M_i)$ and $\delta_{c,i}\equiv\delta_c(z_i)$ for $i=0,1$. The conditions $z_1>z_0$, $S_1>S_0$, and $\delta_{c,1}>\delta_{c,0}$ ensure that a descendant halo of mass $M_0$ at redshift $z_0$ is traced backward in time to its progenitor population at the earlier epoch $z_1$. The conditional first-crossing distribution therefore provides the statistical connection between progenitor and host halos and forms the basis for constructing halo assembly histories and merger trees.\\
In the excursion-set framework, each random-walk trajectory encodes a complete halo assembly history. Consequently, the ensemble of trajectories characterizes the hierarchical growth of dark matter halos and establishes a direct statistical connection between progenitor and descendant populations across cosmic time. These assembly histories play a central role in the present work, providing an important consistency test for the developed merger and accretion model. Specifically, the cumulative mass growth arising from halo mergers and smooth accretion must reproduce the conditional progenitor distribution predicted by Eq.~\eqref{eq:conditional_EST}. Satisfying this requirement ensures that the merger-tree construction remains fully consistent with the underlying excursion-set formalism and accurately captures the statistical properties of hierarchical halo assembly.
	\subsection{Halo Assembly History, Merger and Accretion}\label{sec:mergacc}
				\begin{figure}[t]
		\centering
		\includegraphics[width=\columnwidth]{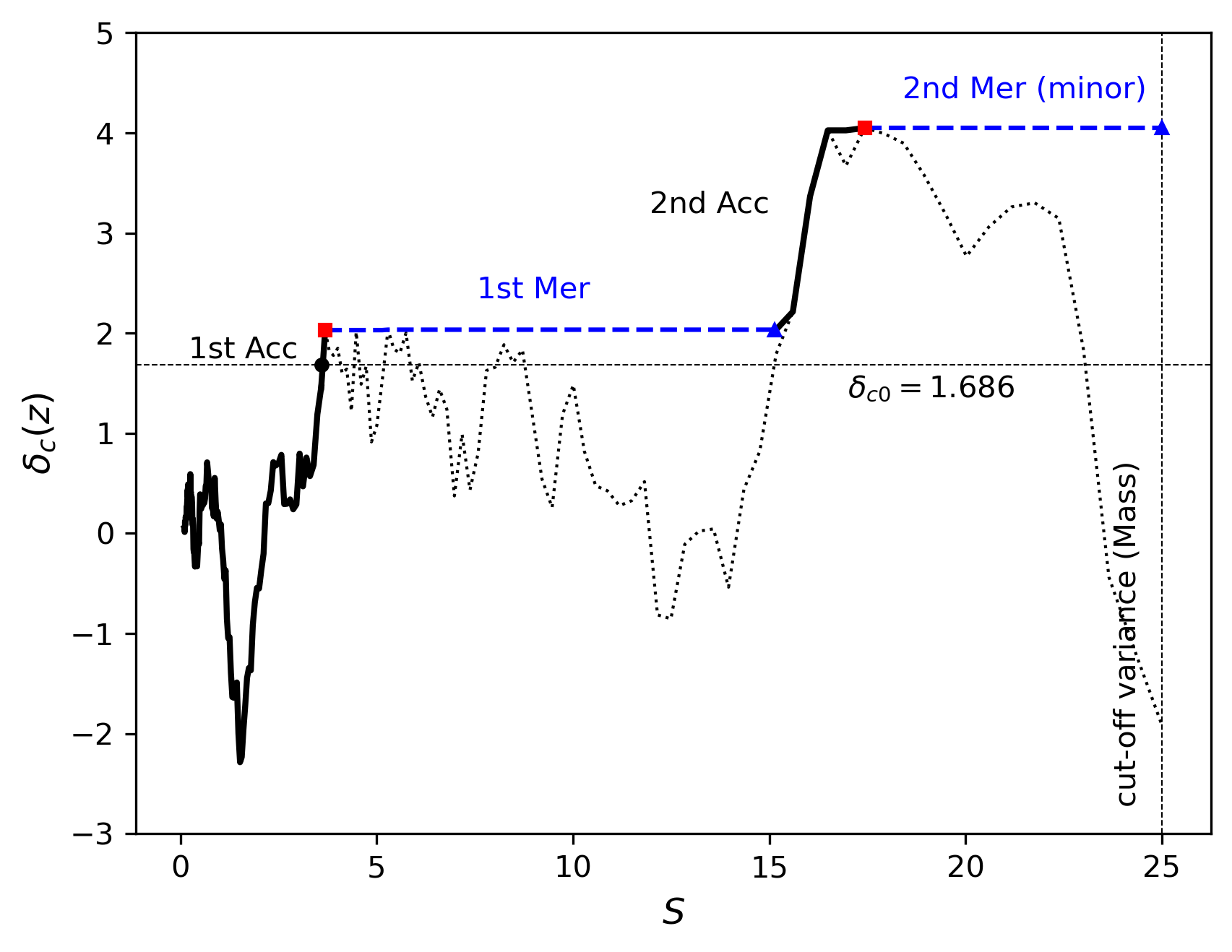}
		\caption{Illustration of the construction of halo assembly histories from Markovian trajectories in EST. The host halo forms at the first up-crossing of the $\delta_{c0}=1.686$. Non-monotonic segments are replaced by horizontal lines to enforce monotonic growth along the main progenitor branch. The resulting assembly history consists of accretion phases (solid black segments) and horizontal jumps at fixed $\delta_c(z)$, which are interpreted as merger events (blue dashed segments). The figure shows two accretion segment (1st Acc and 2nd Acc) and two merger segment (1st Mer and 2nd Mer). The second merger is incorporated into the accretion component because the progenitor mass lies below the resolution threshold, $M_{\rm res}$.}
		\label{fig:fig1}
	\end{figure}
					\begin{figure*}[t]
		\centering
		\includegraphics[width=\textwidth]{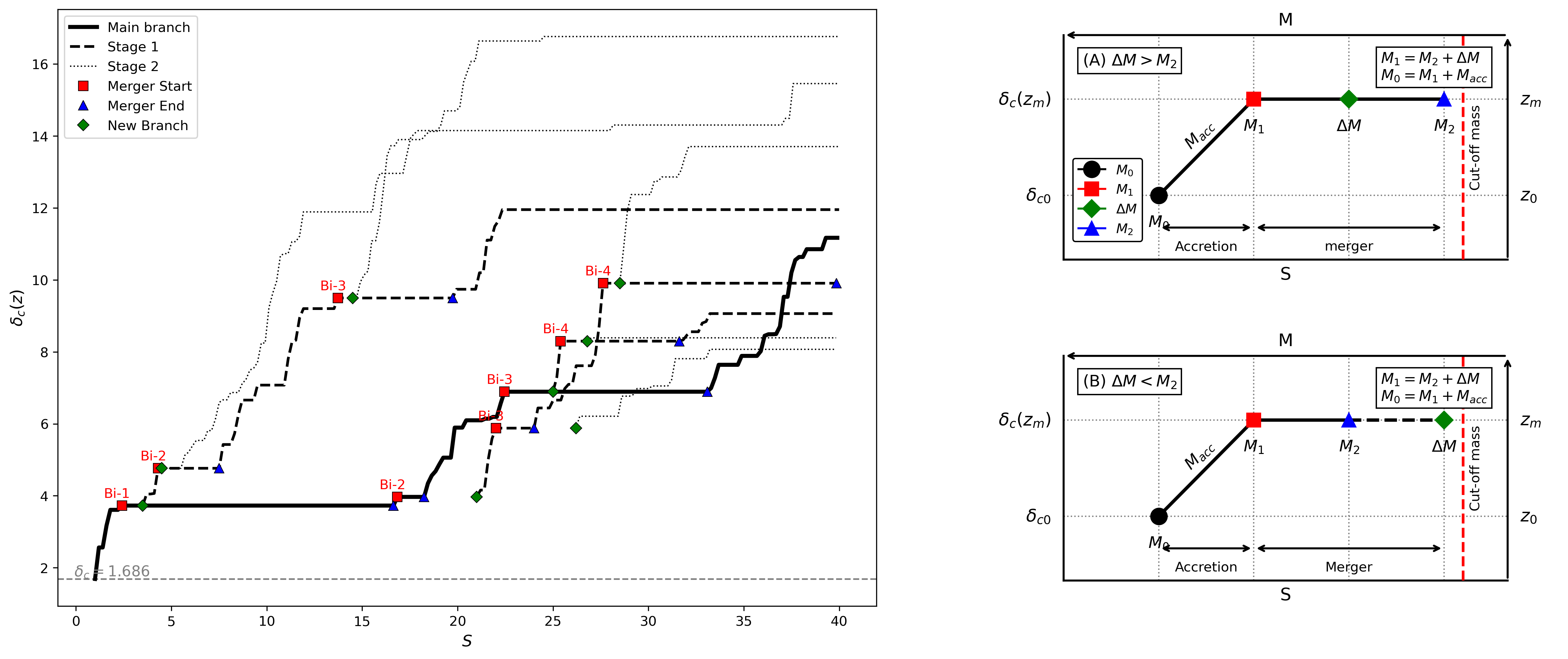}
		\caption{Illustration of the merger-tree construction procedure. Left panel: example merger tree in the $(S,\delta_c)$ plane. The solid black curve denotes the main branch stage, while dashed and dotted curves represent higher-order branch stages. Red squares, blue triangles, and green diamonds indicate merger-start, merger-end, and reconstructed branch points, respectively. Labels Bi-1, Bi-2, etc. identify successive binary level of merger generations.
			Right panel: schematic representation of a single accretion and merger sequence. A halo grows from $M_0$ to $M_1=M_0+M_{\rm acc}$ through smooth accretion. At the merger redshift $z_m$, a descendant halo of mass $M_1$ forms from the merger of two progenitors with masses $M_2$ and $\Delta M$, satisfying $M_1=M_2+\Delta M$. Panels (A) and (B) correspond to the cases $\Delta M>M_2$ and $\Delta M<M_2$, respectively. Recursive application of this procedure generates the full merger tree.}
		\label{fig:fig2}
	\end{figure*}
	These random walk trajectories encode much richer information than the halo abundance statistics described by the simple and conditional mass functions. Within the framework of Excursion Set Theory (EST), a Markovian trajectory originates at the origin of the $(S(M),\delta_c(z))$ plane. When it first crosses the critical density barrier at variance $S_0(M_0)$, a dark matter halo of mass $M_0$ is assumed to form. Subsequent barrier crossings then provide information about the halo mass assembly history and progenitor population at different redshifts. \\
	As the trajectory progresses toward larger values of $S$, corresponding to smaller halo masses, it encounters higher barriers $\delta_c(z)$ at earlier cosmic epochs. Each up-crossing identifies a progenitor halo at the corresponding redshift, allowing the assembly history to be reconstructed from a single trajectory. Halo growth proceeds through two primary processes: smooth accretion, originating from diffuse mass and low-mass structures, and mergers, in which two (or more) progenitor halos combine to form a more massive descendant host halo. \\
	In the hierarchical picture of structure formation, halo mass is expected to grow monotonically with time. To enforce this condition, we impose monotonic growth on the random-walk trajectory. As illustrated in Fig.~\ref{fig:fig1}, once a trajectory reaches a local maximum, corresponding to the formation of a progenitor halo at a given redshift, any subsequent decrease below this value is considered inconsistent with monotonic growth. These segments are replaced by horizontal lines at the local maximum. This procedure naturally generates finite horizontal jumps at fixed $\delta_c(z)$, which we interpret as merger events. Since each jump occurs at a single redshift and connects two distinct values of $S$, it represents the assembly of a larger halo through the merger of smaller progenitors. The endpoint of the jump corresponds to the merging progenitor, while the starting point represents the host halo. Because only one random-walk realization is followed, the resulting history traces only the main branch of the merger tree. \\
	After the trajectory returns to its previous maximum, it resumes monotonic growth and crosses barriers corresponding to progressively higher redshifts. These continuous segments represent smooth accretion. The assembly history therefore consists of an alternating sequence of accretion phases and merger events. This evolution is followed down to a cut-off mass, $M_{\rm res}$, below which progenitors are not individually resolved and their mass contribution is incorporated into the accretion component. \\
	An example of an assembly history is shown in Fig.~\ref{fig:fig1}. The trajectory first up-crosses the barrier $\delta_{c,0}$ at $S \simeq 4$, corresponding to the formation of the host halo at $z=0$. It then evolves to approximately $(S,\delta_c)\simeq(4.2,2)$, corresponding to $z\simeq0.3$, representing the first accretion phase "1st Acc". The subsequent horizontal segment "1st Mer" is identified as a merger event, in which a progenitor with variance $S\simeq15$ merges into a descendant halo with variance $S\simeq4.2$. The following segments "2nd Acc" and "2nd Mer" similarly denote a later accretion phase and merger event. \\
	The second merger is not classified as merger because the merging progenitor has a mass below the adopted resolution threshold, $M_{\rm res}$, and is therefore included in the accretion mass. In general, each random-walk realization generates a unique sequence of accretion and merger events, leading to a distinct halo assembly history. These trajectories provide a physically motivated way of reconstructing halo assembly histories and the hierarchical growth of cosmic structure within the EST framework.\\
	\section{Tree Generation Algorithm}
	By generating a large ensemble of trajectories and reconstructing their corresponding halo assembly histories, it is possible to statistically characterize the hierarchical growth of structure and quantify merger rates, accretion rates, and the relative contributions of mergers and accretion to halo mass assembly. However, the assembly history extracted from a single trajectory corresponds only to the main branch of a merger tree. As discussed in the previous subsection, each merger event identified along the main branch provides information about only one progenitor halo and its descendant. The second progenitor and its subsequent assembly history are not contained in the trajectory itself. A complete description of halo formation therefore requires the recursive identification of all merger counterparts and the construction of the full merger tree. In this subsection, we describe the merger-tree algorithm adopted in this work.\\
	Merger trees are commonly generated using two different approaches. The first and most widely used method is based on discrete time steps. In this framework, specially by the GALFORM binary merger-tree algorithm \cite{Cole2000}, the merger history of a halo is reconstructed backward in time using sufficiently small redshift intervals such that at most one binary merger occurs within each time step. The conditional first-crossing distribution is then used and compared with a uniform random variable to probabilistically determine whether halo growth proceeds through accretion or merger. Variants of this approach include non-binary merger trees \cite{somerville1999plant}, algorithms calibrated against N-body simulations \cite{cole2008statistical,Parkinson2008}, and extensions based on ellipsoidal collapse \cite{Zhang2008}. A comprehensive comparison of these methods is provided by Jiang et al.~\cite{jiang2014generating}.\\
	The second approach, adopted in present work, was introduced by Moreno et al.~\cite{moreno2008merger,moreno2009dark}. Instead of evolving the system through discrete time intervals, the merger tree is constructed through discrete intervals in variance $S$. This formulation is naturally linked to the random-walk trajectories of EST and therefore remains self-consistent with the underlying stochastic process and the conditional first up-crossing distribution. Moreno et al. demonstrated good agreement between this method and N-body simulations and showed that it can be generalized to ellipsoidal-collapse models.\\
		\begin{figure}[t]
		\centering
		\includegraphics[width=\columnwidth]{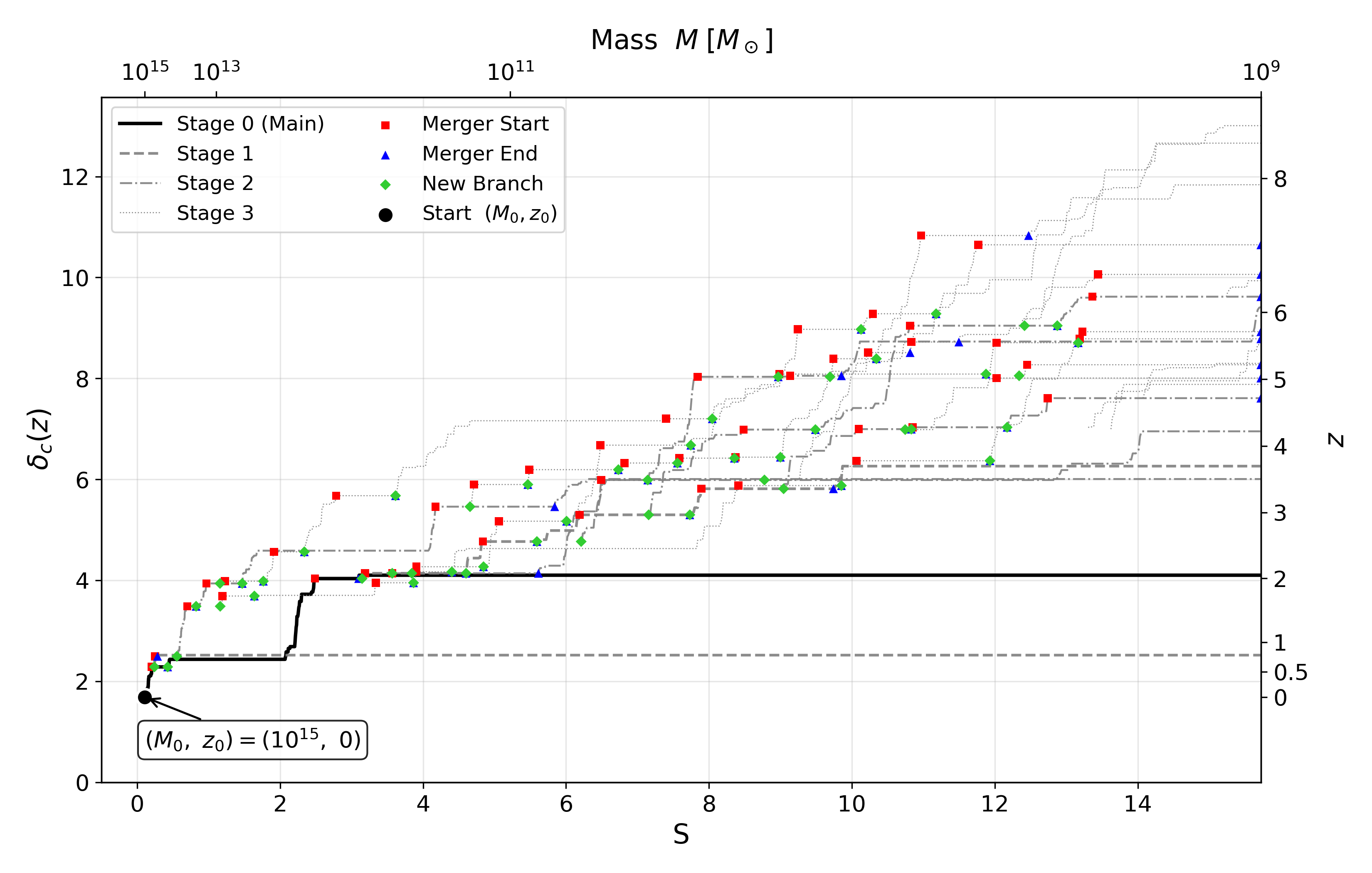}
		\caption{Example merger tree for a host halo of mass $M_0=10^{15},M_\odot$ at $z_0=0$. The tree is displayed in the $(S,\delta_c)$ plane, with the corresponding halo mass and redshift shown on the upper and right axes. The main branch (Stage 0) and three higher-order branching stages are distinguished by different line styles. Horizontal segments represent mergers and rising segments correspond to smooth accretion. Merger-start, merger-end, and reconstructed branch points are marked by red squares, blue triangles, and green diamonds, respectively. The black circle indicates the host halo at $(M_0,z_0)$.}
		\label{fig:fig3}
	\end{figure}
	We adopt the Moreno et al. framework with two modifications. First, we recursively reconstruct all merger branches without enforcing the principal branch to remain the dominant branch throughout the tree construction. Second, we introduce a merger-ratio criterion that supplements the conventional residual-mass threshold and provides a more physically motivated distinction between mergers and accretion.\\
	We begin by specifying a host halo of mass $M_0$ at redshift $z_0$, corresponding to variance $S_0$ and collapse threshold $\delta_{c,0}$. A Markovian trajectory is then generated in the $(S,\delta_c)$ plane, as illustrated in Fig.~\ref{fig:fig1}. Halo mass assembly is extracted from the trajectory by identifying horizontal segments following local maxima as merger events and monotonic segments as smooth accretion. This procedure determines only the main branch of the merger tree.\\
	To reconstruct the complete tree, each merger event is assumed to satisfy binary mass conservation,
	\begin{equation}
		M_1=M_2+\Delta M,
	\end{equation}
	where $M_1$ is the descendant halo mass and $M_2$ and $\Delta M$ are the two progenitor masses. One progenitor corresponds directly to the endpoint of the horizontal segment, while the second progenitor is reconstructed from
	\begin{equation}
		\Delta M(S_{\rm New\,Branch})
		=
		M_1(S_{\rm Merger\,Start})
		-
		M_2(S_{\rm Merger\,End}).
	\end{equation}
	The corresponding variance is obtained from the mass-variance relation, assuming a standard $\Lambda$CDM cosmology with Planck 2018 parameters \cite{planck2020planck}. A new random-walk trajectory is then generated from this point, providing the assembly history of the second progenitor. The procedure is applied recursively to all newly generated branches. We find that the statistical properties of the merger trees converge rapidly, with differences below $1\%$ between trees generated using 10 and 100 branching stages. Throughout this work, we therefore adopt 10 branching stages.\\
	Figure~\ref{fig:fig2} illustrates the recursive construction of the merger tree. A branch begins with halo mass $M_0$ at redshift $z_0$ and grows through smooth accretion by an amount $M_{\rm Acc}$. At redshift $z_m$, the halo reaches mass $M_1$ and undergoes a merger, producing two progenitors with masses $M_2$ and $\Delta M$. These progenitors subsequently evolve along their own assembly histories and may themselves experience further sequence of accretion and merger. Repeating this procedure recursively generates the complete merger tree.\\	
	Two distinct merger configurations can arise (see right panel of Fig. \ref{fig:fig2}). If $\Delta M > M_2$, the reconstructed branch lies between the $M_1$ and $M_2$ locations in the $(S,\delta_c)$ plane. Conversely, if $\Delta M < M_2$, the new branch appears at larger variance than $M_2$. An example merger tree generated for a host halo of mass $M_0=10^{15}\,M_\odot$ at $z=0$ is shown in Fig.~\ref{fig:fig3}. The hierarchical growth of the system can be followed from low-mass progenitors at high redshift toward the final host halo at the present epoch.\\
	It is important to emphasize that the branching stage does not correspond directly to cosmic time or halo mass. A merger event occurring in a higher stage may nevertheless correspond to an earlier epoch or involve more massive halos than events belonging to lower stages. Hence, to further characterize the merger hierarchy, we introduce a binary-level classification. Merger events are labeled as Bi-1, Bi-2, and so forth according to their progenitor generation. Bi-1 events correspond to mergers directly associated with the host halo, while Bi-2 events represent mergers involving their progenitors. This classification facilitates the analysis of merger statistics at different hierarchical binary levels. We note that the resulting trees are not strictly binary. Instead, they are semi-binary, since non-binary configurations arise naturally whenever the accreted mass between two merger events vanishes, $M_{\rm Acc}=0$.\\
	A second modification relative to the original Moreno et al. algorithm concerns the treatment of low-mass mergers. Moreno et al. adopted only a residual-mass threshold, typically $10^{-3}$ of the host-halo mass, below which all contributions are classified as accretion. In contrast, we additionally define a merger ratio,
	\begin{equation}
		MR=
		\begin{cases}
			\dfrac{M_2}{\Delta M}, & \Delta M>M_2,\\[6pt]
			\dfrac{\Delta M}{M_2}, & \Delta M<M_2,
		\end{cases}
	\end{equation}
	where $MR\le1$ by construction.\\
	Only events exceeding a specified minimum merger ratio are classified as mergers; smaller-ratio events are incorporated into the accretion component. Consequently, our algorithm combines both a residual-mass threshold and a merger-ratio criterion. This enables merger trees to be followed to substantially smaller masses, typically down to $10^{-9}M_0$, while preserving physically meaningful merger information. For example, a merger involving halos of masses $10^7\,M_\odot$ and $10^5\,M_\odot$ corresponds to $MR=0.01$. Such an event can still be retained in our framework, whereas conventional residual-mass prescriptions would classify it entirely as accretion once the progenitor masses fall below the adopted residual mass.\\
	We contend that this approach is more physically motivated because the distinction between mergers and accretion should depend primarily on the relative masses of the progenitors rather than on an absolute mass threshold. Moreover, the merger-ratio criterion naturally enables the study of major-merger populations, such as systems with $MR\ge0.25$, which are particularly relevant for observational studies and galaxy-formation models.\\
\begin{figure*}[t]
	\centering
	\includegraphics[width=\textwidth]{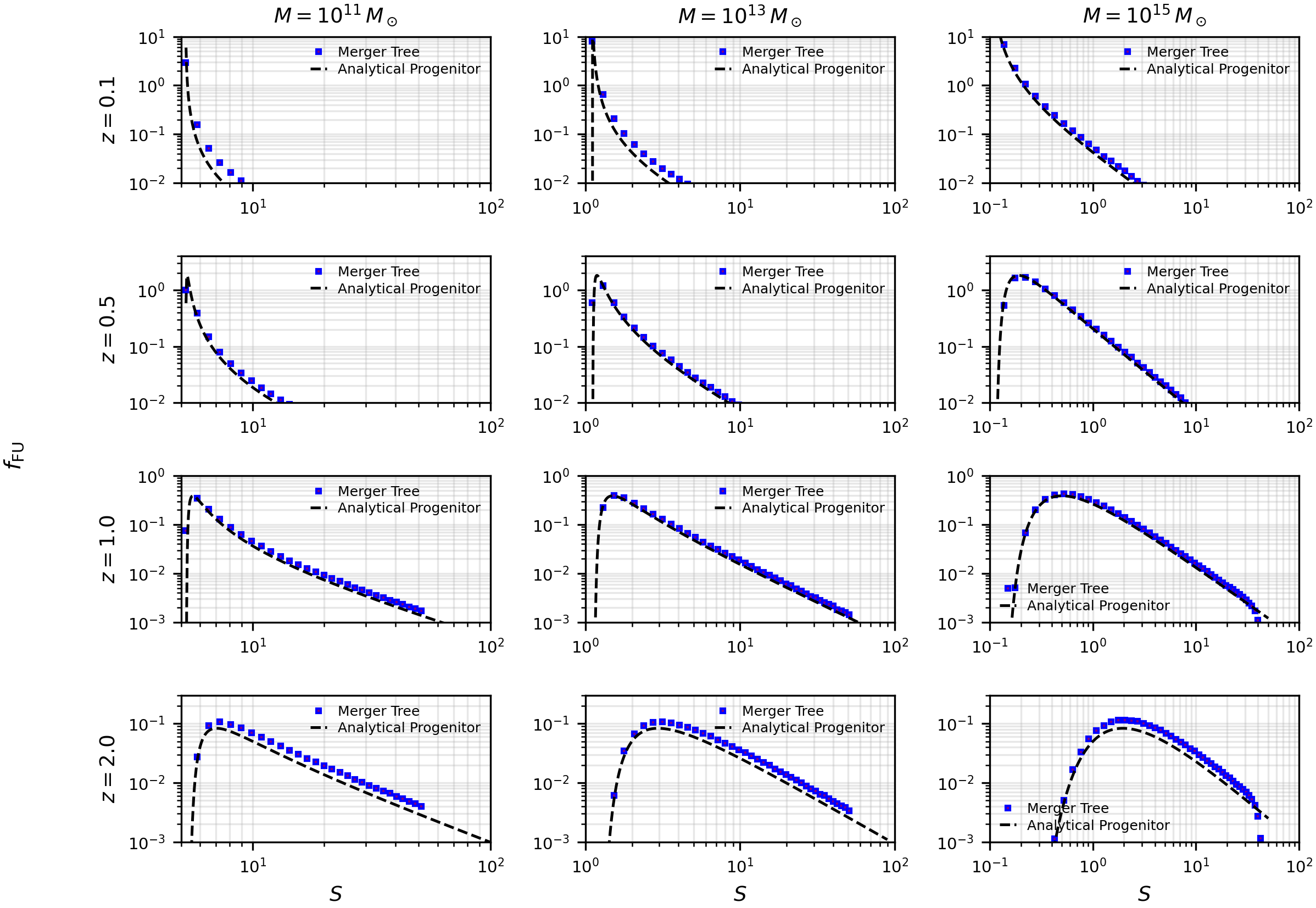}
	\caption{		Consistency test of the merger-tree algorithm. Black dashed curves show the analytical conditional first up-crossing distribution (Eq.~\ref{eq:conditional_EST}), while blue squares denote the progenitor mass distributions obtained from $10^5$ merger-tree realizations. Rows correspond to progenitor redshifts $z=[0.1,0.5,1.0,2.0]$ and columns to host halo masses $M_0=[10^{11},10^{13},10^{15}],M_\odot$ at $z_0=0$. The excellent agreement demonstrates that the merger-tree realizations faithfully reproduce the analytical statistics of excursion-set theory.}
	\label{fig:fig4}
\end{figure*}
	\section{Results and Discussion }\label{sec:results}
	\subsection{Consistency Test}\label{sec:consistency}
	The most direct consistency test of a merger-tree algorithm is its ability to reproduce the analytical conditional progenitor mass function, or equivalently the conditional first up-crossing distribution. This quantity specifies the distribution of progenitor halo masses at redshift $z$ that assemble into a host halo of mass $M_0$ at redshift $z_0$ (Eq.~\ref{eq:conditional_EST}) and naturally incorporates the combined effects of mergers and smooth accretion. Consequently, a statistically complete ensemble of merger-tree realizations should recover the analytical prediction.\\
	More specifically, sampling all progenitor masses between $M_0$ and $M_1$ (corresponding to each sequences accretion-merger) at a fixed redshift should reproduce the conditional progenitor mass function connecting the host halo at $z_0$ to its progenitor population at redshift $z$. This test provides one of the most fundamental validations of a merger-tree algorithm and has been extensively employed in previous studies \cite{Cole2000,Parkinson2008,cole2008statistical,moreno2008merger,moreno2009dark,jiang2014generating}.\\
	Since the present method is based on stochastic realizations of trajectories, consistency with the analytical progenitor mass function is expected by construction. To verify this explicitly, we compare the merger-tree predictions with the analytical result for host halo masses $M_0=[10^{11},10^{13},10^{15}]\,M_\odot$ at $z_0=0$, and evaluate the progenitor mass distributions at redshifts $z=[0.1,0.5,1.0,2.0]$. The results are shown in Fig.~\ref{fig:fig4}, where the analytical prediction is represented by the black dashed curves and the merger-tree results by blue square symbols.\\
	For all plots, we adopt a residual-mass threshold of $M_{\rm res}=M_0/10^9$ and a merger-ratio threshold of $MR=10^{-9}$, allowing all merger and accretion events down to the resolution limit to be resolved. The mass range (variance) is sampled using 1000 bins, and each progenitor distribution is constructed from $N_{\rm re}=10^5$ independent merger-tree realizations.\\
	The progenitor mass functions shown in Fig.~\ref{fig:fig4} are computed using only the main branch of each realization. This choice ensures a one-to-one correspondence between progenitor halos at redshift $z$ and the analytical conditional progenitor mass function. Including higher-order branches would lead to multiple counting of the same mass elements through successive merger generations and would therefore bias the comparison.\\
	As shown in Fig.~\ref{fig:fig4}, the merger-tree results are in excellent agreement with the analytical conditional progenitor mass function over the wide range of halo masses and redshifts considered. The agreement demonstrates that the merger-tree realizations faithfully reproduce the analytical statistics of excursion-set theory and correctly capture the combined effects of mergers and smooth accretion during halo assembly. A more detailed comparison with N-body simulations can be found in the original studies of the Moreno et al. algorithm \cite{moreno2008merger,moreno2009dark}.\\
	\subsection{Accretion Results}\label{sec:accretion}
			\begin{figure}[t]
		\centering
		\includegraphics[width=\columnwidth]{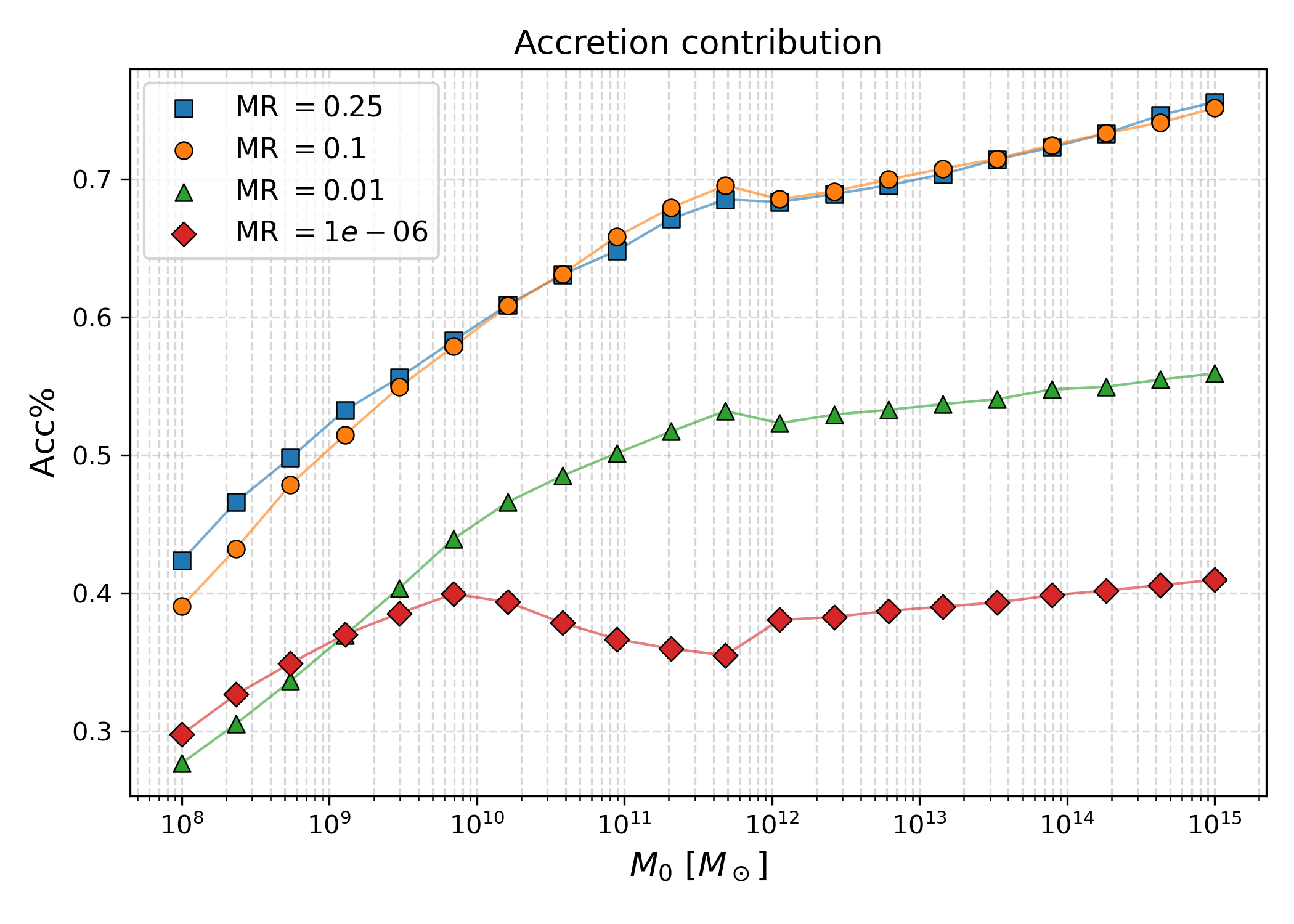}
		\caption{The total contribution of accretion to the formation of a halo as a function of halo mass $M_0$, for different merger ratios $\mathrm{MR} = 0.25, 0.1, 0.01, 10^{-6}$. As the merger ratio is reduced, more and more minor mergers are resolved as mergers rather than counted as accretion, so the accreted fraction decreases.}
		\label{fig:fig7}
	\end{figure}
	\begin{figure*}[t]
		\centering
		\includegraphics[width=\textwidth]{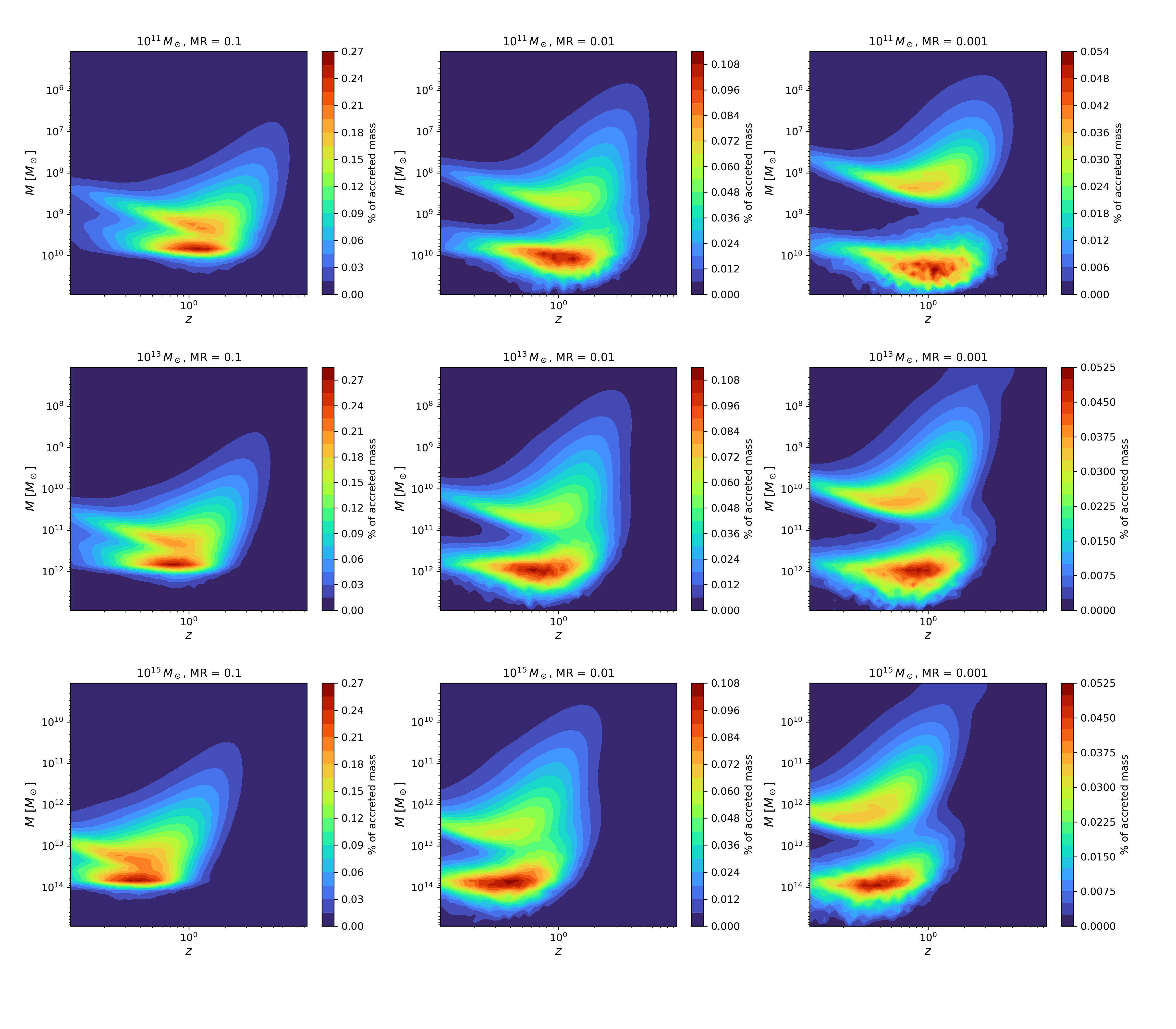}
		\caption{Contour plots illustrating the distribution of accreted mass as a function of halo mass (vertical axis) and redshift (horizontal axis). The color scale indicates the percentage of the host halo mass contributed by accretion within each ($M$, $z$) bin. The panels are organized by host halo mass ($M = 10^{11}, 10^{13}, 10^{15}\ M_\odot$; rows) and mass ratio ($\mathrm{MR} = 0.1, 0.01, 0.001$; columns). For a fixed MR, the distribution remains nearly self-similar across different halo masses, differing only by a proportional shift in the mass range. However, for a given host halo mass, decreasing the MR results in a broader distribution of accretion across the $M$--$z$ plane.}
		\label{fig:fig5}
	\end{figure*}
	\begin{figure*}[t]
		\centering
		\includegraphics[width=\textwidth]{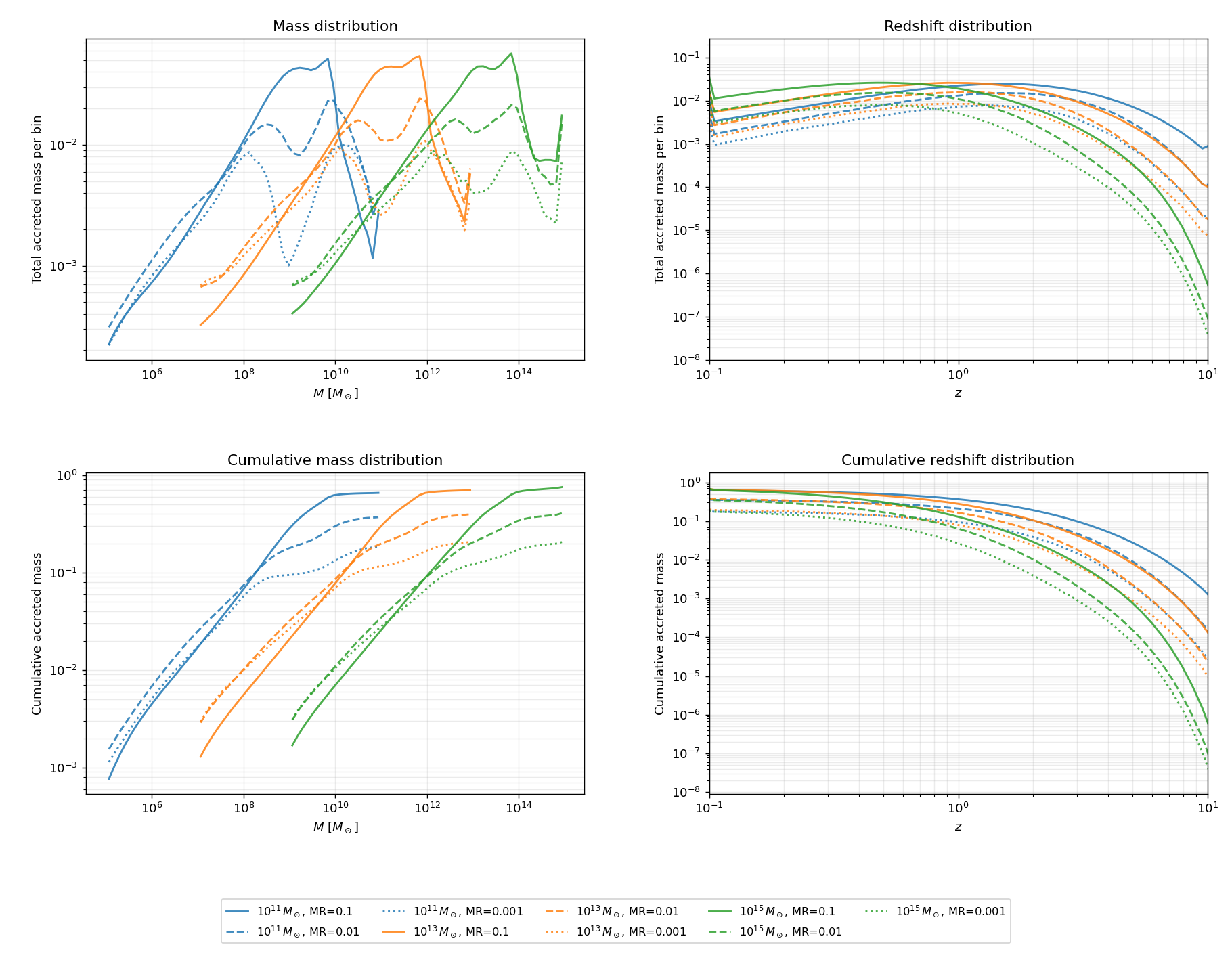}
		\caption{The distribution of accreted mass as a function of halo mass (left panels) and redshift (right panels). The upper panels show the differential distribution per mass/redshift bin, while the lower panels display the corresponding cumulative distributions. Results are presented for different host halo masses, where colors represent $M = 10^{11}, 10^{13}, 10^{15}\ M_\odot$ and line styles indicate mass ratios ($\mathrm{MR} = 0.1, 0.01, 0.001$).}
		\label{fig:fig6}
	\end{figure*}
	Genel et al.~\cite{genel2010growth} used merger trees extracted from the Millennium and Millennium-II simulations to quantify the relative contributions of mergers and smooth accretion to dark matter halo growth. They found that mergers with mass ratios larger than $3{:}1$ and $10{:}1$ contribute approximately $20\%$ and $30\%$ of the final halo mass, respectively, while all resolved mergers combined account for roughly $60\%$. Applying our EST-based merger-tree algorithm to the same problem, we recover these results with excellent agreement, demonstrating that the excursion-set formalism successfully reproduces the merger and accretion contributions inferred from large cosmological N-body simulations.\\
	The computational efficiency of the EST framework allows us to extend this analysis over a broad range of halo masses and redshifts. Figure~\ref{fig:fig7} shows the accretion contribution as a function of host halo mass for merger-ratio thresholds $MR=\{10^{-6},\,0.01,\,0.1,\,0.25\}$ over the host halo mass range $10^{8}\lesssim M_0/M_\odot\lesssim10^{15}$. Each merger tree is evolved through 10 stages, and the statistical analysis is based on $N_{\rm re}=5\times10^{5}$ realizations of trajectories. The merger ratio defines the minimum progenitor mass ratio classified as a merger, while smaller ratios are incorporated into the accretion component.\\
	Several important outcome emerge. First, the accretion contribution increases systematically with halo mass for all merger-ratio thresholds. Second, the inferred accretion fraction depends strongly on the adopted merger-ratio criterion. For major-merger thresholds ($MR=0.1$ and $MR=0.25$), the accretion fraction reaches approximately $70\%$ for massive halos, implying that major mergers contribute only $\sim30\%$ of the final halo mass. This result is fully consistent with the analysis of Genel et al.~\cite{genel2010growth}, who likewise concluded that major mergers account for only a subdominant fraction of halo growth.\\
	At the opposite extreme, the threshold $MR=10^{-6}$ effectively includes all resolved mergers down to the adopted mass resolution. In this case, the accretion fraction decreases to approximately $40\%$, implying that mergers contribute roughly $60\%$ of the final halo mass. Once again, the agreement with the Millennium simulation results of Genel et al.~\cite{genel2010growth} is remarkably good. The smooth variation of the accretion fraction with merger ratio further supports the robustness of the merger-accretion classification adopted in this work. 	Overall, the close agreement between the EST predictions and high-resolution N-body simulations demonstrates that the semi-analytical framework accurately captures the principal channels of hierarchical halo growth while remaining computationally inexpensive.\\
	Figure~\ref{fig:fig7} provides the cumulative contribution of smooth accretion integrated over the entire assembly history of each halo. Although these results quantify the overall importance of accretion, they do not reveal when the accreted mass is acquired or which progenitor populations dominate the growth process. To address these questions, we examine the distribution of accreted mass in the mass-redshift plane.\\
	Figure~\ref{fig:fig5} presents accretion-distribution contours for host halo masses $M_0=\{10^{11},10^{13},10^{15}\}\,M_\odot$ and merger-ratio thresholds $MR=\{0.001,0.01,0.1\}$. The color scale represents the fractional contribution of accretion within each $(M,z)$ bin. By construction, integration over the entire distribution yields the total accretion fraction shown in Fig.~\ref{fig:fig7}.\\
	Several remarkable features are immediately apparent. First, for a fixed merger ratio, the contour morphology remains nearly unchanged across the full range of halo masses. The distributions differ primarily by a shift along the mass axis proportional to the final halo mass, indicating a nearly self-similar accretion process. The dominant contribution consistently originates from progenitors in the range
	\begin{equation}
		10^{-3}M_0 \lesssim M \lesssim 10^{-1}M_0,
	\end{equation}
	suggesting that a substantial fraction of smooth halo growth is associated with progenitors containing roughly $0.1\%-10\%$ of the final halo mass, almost independently of host-halo mass.\\
	A second important feature concerns the redshift dependence of accretion. In all cases, the largest contributions are concentrated within the interval
	\begin{equation}
		0.5 \lesssim z \lesssim 2,
	\end{equation}
	while the contribution declines rapidly at earlier epochs. This indicates that smooth accretion is predominantly a low- to intermediate-redshift process. The location of the peak contribution also shifts systematically with halo mass. For $M_0=10^{11}\,M_\odot$, the maximum occurs near $(z,M)\simeq(1,10^{10}\,M_\odot)$, whereas for $M_0=10^{15}\,M_\odot$ it moves toward lower redshift and larger progenitor mass. This trend implies that increasingly massive halos continue assembling a significant fraction of their mass at later cosmic times.\\
	Interestingly, the location of the dominant accretion region depends only weakly on the merger-ratio threshold. Lowering $MR$ increases the total accretion fraction by incorporating progressively smaller progenitors into the accretion component, but it does not significantly alter the characteristic progenitor masses or redshifts responsible for the bulk of the growth.\\
	To further quantify these trends, Fig.~\ref{fig:fig6} shows the one-dimensional mass and redshift projections of the contour distributions. The differential mass distributions exhibit nearly identical shapes after accounting for the overall scaling with host-halo mass, providing additional evidence for the self-similar nature of the accretion process. In all cases, the peak contribution occurs at progenitor masses of approximately $10^{-3}$--$10^{-1}$ of the final halo mass, consistent with the contour analysis.\\
	The redshift projections reveal that the majority of accretion occurs at low and intermediate redshifts, with the strongest contribution arising within $0.5\lesssim z \lesssim2$. More massive halos retain a larger accretion contribution at late times, confirming the tendency of massive halos to continue accreting matter at late times.\\
	The cumulative distributions provide an integrated view of halo growth. Both the mass and redshift cumulative functions exhibit remarkably similar shapes across different halo masses and merger-ratio thresholds, reinforcing the universal character of the accretion process. Together, these results indicate that smooth accretion follows a nearly self-similar mass distribution, dominated by progenitors within $10^{-3}M_0\lesssim M\lesssim10^{-1}M_0$, while its temporal evolution reflects the hierarchical nature of structure formation, with increasingly massive halos assembling a substantial fraction of their mass at comparatively recent epochs.\\
	\section{Conclusion}\label{sec:conclusion}
	In this work, we investigated the relative contributions of mergers and smooth accretion to the assembly of dark matter haloes within the framework of EST. We adopted the variance-discretized merger-tree algorithm developed by Moreno et al.\cite{moreno2008merger,moreno2009dark} and extended it by introducing a merger-ratio criterion in addition to the conventional residual-mass threshold. This modification provides a physically motivated distinction between mergers and accretion and allows halo growth to be studied as a function of merger ratio.\\
	The reliability of the merger-tree algorithm was verified by comparing the progenitor mass function obtained from the merger trees with the analytical conditional first up-crossing distribution of EST. Excellent agreement was obtained over a broad range of halo masses and redshifts, demonstrating that the algorithm accurately reproduces the statistical predictions of the underlying excursion-set formalism.\\
	Applying this framework to halo assembly histories, we found that the inferred accretion contribution depends strongly on the adopted merger-ratio threshold. For large merger ratio ($MR=0.1$ and $MR=0.25$), smooth accretion contributes up to approximately $70\%$ of the mass growth of massive haloes, implying that mergers account for only about $30\%$ of their final mass. In contrast, when all resolved mergers are included ($MR=10^{-6}$), the accretion contribution decreases to approximately $40\%$, corresponding to a merger contribution of roughly $60\%$. These results are in very good agreement with the N-body simulation analysis of Genel et al.\cite{genel2010growth}, providing strong support for the consistency of the EST-based approach.\\
	We further investigated the origin of the accreted mass using its distribution in the mass-redshift contour. The results reveal a nearly self-similar accretion pattern across a wide range of host halo masses when progenitor masses are expressed relative to the final halo mass. The dominant contribution to smooth accretion originates from progenitors in the range $10^{-3}M_0 \lesssim M \lesssim 10^{-1}M_0$, with the differential distributions typically peaking near $10^{-2}$--$10^{-1}M_0$. The accretion process is concentrated primarily within the redshift interval $0.5 \lesssim z \lesssim 2$, while the peak accretion epoch shifts toward lower redshift for increasingly massive host haloes. These results indicate that massive haloes continue assembling a significant fraction of their mass at comparatively recent cosmic times.\\
	The proposed EST-based framework provides an efficient alternative to large cosmological N-body simulations for studying merger histories, accretion processes, and halo assembly over a broad range of masses, redshifts, and merger definitions. In addition, a GPU-based implementation developed for this work accelerates the calculations by up to two orders of magnitude relative to the original CPU implementation, enabling the generation of very large statistical ensembles of merger trees.\\
	The resulting merger-tree framework provides a flexible and computationally efficient tool for studying halo assembly, galaxy formation, and the evolution of cosmic structures. By separating halo growth into major mergers, minor mergers, and smooth accretion, the method enables investigations of the connection between halo assembly histories, internal halo properties, galaxy evolution, and cosmological models \cite{lee2014sussing,wang2011assembly,correa2015accretion}. The present study is based on Markovian trajectories, a spherical-collapse barrier, and the standard $\Lambda$CDM cosmology. Future extensions to ellipsoidal-collapse models, non-Markovian trajectories, alternative smoothing filters, and different cosmological scenarios will provide a powerful framework for exploring the influence of cosmology and dark-matter physics on the hierarchical growth of structure.\\
\begin{acknowledgments}
The author acknowledges the use of Microsoft Copilot, ChatGPT, and DeepSeek for assistance with language editing, text rephrasing, and improving the clarity and academic presentation of the manuscript. These tools were used as aids during manuscript preparation and did not contribute to the scientific interpretation, methodology, and analysis of this work
\end{acknowledgments}
	\bibliography{references}
\end{document}